\documentclass[reprint, prl]{revtex4-2}
\usepackage{amssymb}
\usepackage{amsmath}
\usepackage{braket}
\usepackage{color}
\usepackage{color}
\usepackage{comment}
\usepackage{graphicx}
\usepackage{hyperref}

\newcommand{\be}{\begin{equation}}
\newcommand{\ee}{\end{equation}}
\newcommand{\ba}{\begin{aligned}}
\newcommand{\ea}{\end{aligned}}
\newcommand{\bs}[1]{\boldsymbol{\bf #1}}

\newcommand{\HS}{\chi}

\newcommand{\Z}{z}

\newcommand{\tr}{\mathrm{tr}}

\begin{document}

\title{
Macroscopic Quantum States and Universal Correlations\\in a 
Disorder-Order Interface Propagating over a 1D Ground State
}
\author{Vanja Mari\'c}
\author{Florent Ferro}
\author{Maurizio Fagotti}
\affiliation{Universit\'e Paris-Saclay, CNRS, LPTMS, 91405, Orsay, France}

\date{\today}

\begin{abstract}
We consider translationally invariant quantum spin-$\frac{1}{2}$ chains with local interactions and a discrete symmetry that is spontaneously broken at zero temperature. We envision  experimenters switching off the couplings between two parts of the system and preparing them in independent equilibrium states. One side of the chain {is prepared in a disordered phase, and the other in a }symmetry-breaking ground state. When the couplings are switched back on, time evolution ensues. We argue that in integrable systems the front separating the ordered region recedes at the maximal velocity of quasiparticle excitations over the ground state. We infer that, generically, the order parameters should vary on a subdiffusive scale of order $t^{1/3}$, where 
$t$ is time, and their fluctuations should exhibit the same scaling. {This} interfacial region 
 exhibits full range correlations, indicating that it cannot be decomposed into nearly uncorrelated subsystems. 
Using the transverse-field Ising chain as a case study, we demonstrate that all order parameters follow the same universal scaling functions. 
Through an analysis of the skew information, we uncover that the breakdown of cluster decomposition has a quantum contribution: each subsystem within the interfacial region, with extent comparable to the region, exists in a macroscopic quantum state.  
\end{abstract}

\maketitle

Local relaxation is a key concept in the study of isolated quantum many-body systems prepared out of equilibrium~\cite{Polkovnikov2011Colloquium,Gogolin2016Equilibration}. One of the key aspects of
their late-time description  is
indeed that, eventually, small parts of the system behave as if they were in thermal equilibrium, which is often explained invoking the eigenstate thermalization hypothesis~\cite{Rigol2008, deutsch91,Srednicki1994Chaos,Deutsch_2018}. Exception to this local form of thermalization are well known in 1D, where integrable and quasi-integrable systems stand out for their peculiarities,  observed even in experiments~\cite{Kinoshita2006,Hofferberth2007,Gring2012,Langen2015}.
In conventional situations also integrable systems exhibit local relaxation, but the expectation values of local observables are not captured by Gibbs ensembles whereas by so-called generalised Gibbs ensembles (GGE)~\cite{Rigol2007Relaxation,Doyon2017Thermalization,Vidmar2016Generalized,Essler2016Quench}, which account for the presence of infinitely many integrals of motion.
In integrable systems with inhomogeneities, local relaxation was shown in a multitude of works, culminating in the theory of generalized hydrodynamics (GHD)~\cite{Castro-Alvaredo2016Emergent,Bertini2016Transport,Borsi2020,Alba2021Generalized,DeNardis2022Correlation,Bulchandani2021Superdiffusion,Borsi2021Current,Essler2023review}, whose deviations from standard hydrodynamic predictions have been {detected} also in experiments~\cite{Schemmer2019,Malvania2021,Bouchoule2022}.

Note that, since (generalised) Gibbs ensembles are characterised by a nonzero thermodynamic entropy and, in 1D, entropy disrupts order very effectively, every global symmetry common to all relevant integrals of motions is expected to be locally restored~\cite{Fagotti2013Reduced,Essler2016Quench}.
This is supposed to happen everywhere, even beyond the nonequilibrium steady state (NESS) emerging in the genuine infinite time limit;
for instance, in bipartitioning protocols it occurs in the Euler scaling limit in which the position of the subsystem {is regarded as a time-dependent variable.}

We remark, however, that referring to the  emergent stationary states capturing the late-time local properties as to 
states of (generalised) equilibrium might be misleading: stationarity is not sufficient to characterise an equilibrium state~\cite{Haag1974},  which is also supposed to be stable under small perturbations, as well as to exhibit other subtle properties that manifest themselves, e.g., in 
the weak clustering of correlations {(see also Ref.~\cite{Sieberer2015} in the presence of time reversal symmetry)}. In this regard, we mention Ref. \cite{Ogata2004The}, which pointed out a form of instability of the NESS emerging after joining two thermal reservoirs to a finite subsystem.  

In this work, we examine the cluster decomposition properties of the stationary states emerging in late-time descriptions. In an equilibrium state of a local system, experiments conducted at large, space-like separations do not influence each other. In the nonequilibrium time evolution of an inhomogeneous, isolated system---where relaxation is a local phenomenon and subsystems are the focus---it would then be reasonable to expect such a separation not to exceed the maximum size of the subsystem that can be described by a stationary state.
Is that always the case?
Remarkably, we show that even such a weak requirement is not always satisfied.
Fluctuations and, more generally, the full counting statistics of operators in subsystems of nonequilibrium systems has attracted a lot of attention~\cite{Eisler2013,Groha2018,Bastianello2018from,Collura2020How,Myres2020,Bertini2023Nonequilibrium,Senese2023,Zadnik2024}. Several unusual behaviours have been uncovered, including
algebraically decaying correlations~\cite{Doyon2023Emergence,Doyon2023Ballistic,DelVecchioDelVecchio2022,Maric2022Universality,Maric2023universality,Maric2023universality2} and anomalous fluctuations ~\cite{Krajnik2022Absence,Krajnik2022Exact,Krajnik2024,Yoshimura2024anomalouscurrent}.
In those cases it is still not excluded that local stationary states could  be understood as equilibrium states of effective long-range models ({see also Ref.~\cite{Aschbacher2003})}. Instead, we are investigating whether connected correlations might not decay at all in the effective stationary states. One scenario that could lead to this is when time evolution fails to smooth out the discontinuity induced by a bipartite preparation of the system, because there are no clustering stationary states that can bridge the properties of the two sides. In such cases, we could expect an unusually rapid crossover between areas with significantly different properties, even at late times.   If this crossover is sufficiently sharp, the size of the region it influences could become comparable to the magnitude of fluctuations of the extensive observables within that region. 
Since in 1D order can be present only at strictly zero temperature, we focus on the following situation: 
a semi-infinite chain {(with open boundary conditions)} in a symmetry breaking ground state is put in point contact with a complementary semi-infinite chain prepared at higher temperature.

\paragraph{Morphology of the disorder-order interface.}

\begin{figure}[t]
    \centering
    \includegraphics[width=0.47\textwidth]{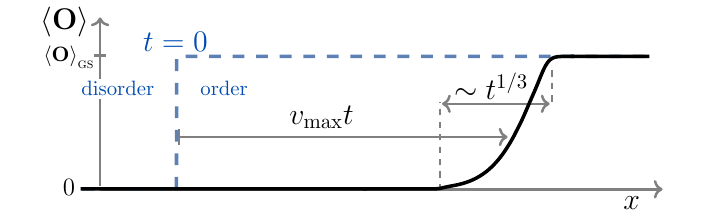}
    \caption{Schematic representation of order depletion, characterized by a local order parameter {$\braket{\bs O}$.}}
    \label{fig:schematic}
\end{figure}

In integrable systems, bipartitioning protocols such as the two-temperature scenario we are considering have been widely investigated~\cite{Bertini2016Transport,Castro-Alvaredo2016Emergent,Piroli2017Transport,Bertini2018Lowtemperature,Bertini2018Universal,Bertini2019Transport}, especially under the framework of generalised hydrodynamics. A qualitative picture has emerged of a lightcone propagating from the junction of the two semi-infinite chains. {The right (left) edge of the lightcone corresponds to the maximal velocity of excitations towards the right (left) over the stationary state describing the late-time behavior of local observables outside the lightcone}. The expectation values of local observables approach stationary values in the Euler scaling limit in which the distance from the junction is proportional to the time, which is sent to infinity. It is customary to call ray the straight curve along which the infinite time limit is taken.

From this perspective, the most visible  effect of interactions is the dependency of the slope of the edges of the lightcone on the states outside the lightcone---see also Ref.~\cite{Bonnes2014Light-Cone}. Such effect becomes even stronger when there are more species of excitations, as they generally exhibit different maximal velocities. Arguably, however, the quintessential difference between noninteracting and interacting systems lies in the diffusive correction to  generalised hydrodynamics, as emphasized in Ref.~\cite{Spohn2018Interacting}. Diffusion is responsible, in particular, of the different scaling behaviour observed at the edges of the lightcone, which is typically $t^{1/2}$~\cite{Collura2018Analytic} rather than the celebrated $t^{1/3}$ of noninteracting systems~\cite{Hunyadi2004,Eisler2013}. To the best of our understanding, however, diffusion in integrable systems with a GHD description is a classical effect, therefore the diffusive contribution should not be expected very close to a symmetry breaking ground state, which is purely quantum in that it lies, isolated (together with the other degenerate ground states), at the edge of the energy spectrum.  This argument  is consistent, for example, with the numerical observations of Ref.~\cite{Zauner2015}, which investigated a bipartitioning protocol in the ordered phase of the XXZ model---an interacting integrable system.  Specifically, they showed that, close to the edge of a lightcone spreading over the ground state, the order parameter approaches its ground state value as $t^{-1/3}$ over a region scaling as $t^{1/3}$, which {resembles} the Tracy-Widom scaling~\cite{Tracy1993Level} typical of noninteracting systems such as the transverse-field Ising chain~\cite{Platini2005Scaling,Perfetto2017Ballistic,Fagotti2017Higher}. 
Thus, we will assume this scaling behavior without significant loss of generality---see also Ref.~\cite{Bulchandani2019Subdiffusive}. Most of the following arguments, however, remain valid even if the expected $t^{1/3}$ is replaced by $t^{\alpha}$, where $0 < \alpha < 1$.

The next important aspect we want to highlight follows from the results of Refs~\cite{Bertini2018Entanglement,Alba2019Entanglement}. Specifically, rare exceptions apart, the entropies of large subsystems  are extensive in the limit of infinite time along any ray strictly inside the lightcone. This is true, in particular, in the two-temperature scenario, even when one side of the system is prepared at zero temperature. In the latter case, however, close to the edge of the lightcone spreading over the ground state, the entropy of subsystems loses its extensive contribution and, in particular, the half chain entropy on the ground-state side becomes finite.
As anticipated when we mentioned symmetry restoration in generalised Gibbs ensembles in 1D, a stationary state with extensive entropy is expected to possess all the symmetries of the local conservation laws. Thus, it is reasonable to expect the absence of order along any ray strictly inside the lightcone. This observation, together with the previous argument supporting the $t^{1/3}$ scaling behaviour, suggests that the order parameter has a tight spatial window, of order $t^{1/3}$, to change from the ground state value to zero. This is the disorder-order interfacial region we are interested in, schematically presented in Fig.~\ref{fig:schematic}. 

{On the other hand, local observables that are invariant under the symmetry that is spontaneously broken retain values close to those of the ground state within the interfacial region (the GHD solution is generally continuous at the edges of the lightcone). There,
}
the effective states describing the local properties at late times
differ from the ground state only by a finite number of excitations. We remind the reader that the quasilocalised packets of single-particle excitations over a symmetry breaking ground state have a domain wall structure interpolating from one ground state to another. 
{While symmetric local observables at sufficiently large distances from those packets remain unaffected, nonsymmetric observables do not. Specifically,}
the value of the order parameter depends on whether a quasilocalised packet is on the left or on the right of the observable, independently of the actual distance.  As long as the density of excitations is zero, {this intrinsic nonlocality}
prevents the connected correlations of order parameters to approach zero within the region, resulting in the breakdown of cluster decomposition. 

{But the story does not end here.}
The previous arguments suggest that, in the limit of infinite time, the state {deviates}
from the ground state only in the correlation functions of nonsymmetric observables, which are expected to 
{vary} on $t^{1/3}$ length scales. In addition, the {interfacial} region is characterised solely by quasiparticles with velocity close to the maximal one---a fraction of order $t^{-1/3}$ of the full momentum space~\footnote{  A quasiparticle with velocity $v(p)$ is in the interfacial region  only if $v(p)t-v_{\textrm{max}}t\sim t^{1/3}$. Assuming that the second derivative of the velocity w.r.t. the momentum is nonzero at the maximum, this leads to the condition $p-\bar p\sim t^{-1/3}$, where $\bar p$ is the momentum that maximizes the velocity.}. 
The interfacial region seems to fulfill all the requirements to be described in a continuum scaling limit, 
which would filter out most of the microscopic details, eventually uncovering the universal character. 

To provide a concrete example, we now prove the qualitative conclusions drawn here using a specific integrable model, which will also help identify additional features to expect in other integrable systems.

\paragraph{The model.}

The transverse-field Ising chain is a paradigmatic model for quantum phase transitions~\cite{Sachdev2011book} and is described by a local spin-$\frac{1}{2}$ Hamiltonian $\bs H_1$, where, for later convenience, we defined
\begin{equation}\label{eq:HIsing}
    \bs H_{s} =-\sum\nolimits_{\ell}[ (1-\delta_{s0}\delta_{\ell 0})\bs\sigma^x_{\ell} \bs\sigma^x_{\ell+1}+h \bs\sigma^z_\ell] \, , 
\end{equation}
$\bs \sigma^\alpha_\ell$ for $\alpha=x,y,z$ act as Pauli matrices on site $\ell$ and as the identity elsewhere; $h$ parametrises the effect of an external magnetic field in the transverse direction.
For $|h|<1$ the model exhibits a zero-temperature ferromagnetic phase in which the spin-flip symmetry associated with the transformation $\bs P=\prod_j\bs\sigma_j^z$ is spontaneously broken. Namely, for $|h|<1$ there are two stable ground states $\ket{\textsc{gs}_\pm}$ with spontaneous magnetization
$
    \braket{\textsc{gs}_\pm|\bs \sigma^x_\ell |\textsc{gs}_\pm}=\pm m^x_{\textsc{gs}}
$, 
where
$
    m^x_{\textsc{gs}}= (1-h^2)^{1/8}
$.
For $|h|>1$, instead, the model exhibits a paramagnetic phase in which spins tend to align in the transverse direction. 

Bipartitioning protocols in this model have been widely investigated~\cite{Aschbacher2003,Platini2005Scaling,Zauner2015,Eisler2016Universal,eisler2020Front,Perfetto2017Ballistic,Perfetto2020Dynamics,Fagotti2017Higher,DeLuca2013Nonequilibrium,Kormos2017,Eisler2018Hydrodynamical,Fagotti2020,Delfino2022Space,Bocini2023Connected}.  
We focus on a protocol 
in which the system is prepared in {an equilibrium} state for the split Ising Hamiltonian $\bs H_0$, with $0<h<1$, in which the coupling between site $0$ and $1$ is turned off.
We denote by $\beta$ the inverse temperature of the left part; the right part is prepared at zero temperature.
The spin-flip symmetry is broken on the right hand side, where  the longitudinal magnetization (far enough from the boundary) is equal to $m_{\textsc{gs}}^x$.
We then consider the local quench consisting in switching on the coupling that  was originally off: $\bs H_0\rightarrow\bs H_1$.

Both before and after the quench the Hamiltonian  is quadratic in the Majorana fermions {$\bs a_{2\ell-1}=(\prod_{j<\ell}\bs \sigma_j^z)\, \bs\sigma_\ell^x$ and $\bs a_{2\ell}=(\prod_{j<\ell}\bs \sigma_j^z)\, \bs\sigma_\ell^y$}---self-adjoint operators that satisfy the algebra
$
    \{\bs a_{\ell}, \bs a_n\}=2\delta_{\ell n}\bs I 
$. This allows one to use free-fermion techniques to work out the expectation value of local observables and entanglement entropies. We remark, however{, that, because of symmetry breaking, the initial state is not Gaussian, i.e. proportional to $\exp{(\sum_{j,\ell} \bs a_j W_{j,\ell} \bs a_\ell})$ for some antisymmetric matrix $W$. This complication} is dealt off with tricks based on clustering properties and Lieb-Robinson bounds~\cite{Lieb1972The,Bravyi2006Lieb}.

Ref.~\cite{Fagotti2020} reformulated time evolution in noninteracting spin chain models such as the Ising one in the framework of phase-space quantum mechanics. 
Specifically, a Gaussian state is completely characterized by a real field, the root density, $\rho_x(p)$, {describing the density of excitations,} and an auxiliary field, $\Psi_x(p)$, which {captures the creation/annihilation of excitations and} is complex and odd under $p\rightarrow -p$. {Variables $(x,p)$ are phase-space coordinates: $p$ is the momentum and $x$ is the  position on the chain (while $x$ is usually extended to $\mathbb R$, physical degrees of freedom correspond only to $x\in\frac{1}{2}\mathbb Z$).} Time evolution is decoupled in  $\rho_x(p)$ and $\Psi_x(p)$, indeed the fields satisfy
\begin{equation}\label{eq:dyneq}
\begin{aligned}
    i \partial_t \rho_{x,t}(p)=&\varepsilon(p)\star\rho_{x,t}(p)-\rho_{x,t}(p)\star\varepsilon(p)\\
    i \partial_t \Psi_{x,t}(p)=&\varepsilon(p)\star \Psi_{x,t}(p)+\Psi_{x,t}(p)\star\varepsilon(-p)\; ,
\end{aligned}
\end{equation} 
where $\varepsilon(p)=2\sqrt{1+h^2-2h\cos p}$ is the energy of the excitation. The operation $\star$ is the Moyal product{, defined in the Appendix}.
We refer the reader to Refs~\cite{Fagotti2020, Alba2021Generalized} for additional details. Here we only mention that the elements of the
correlation matrix {$\Gamma_{j,\ell}=\delta_{j,\ell}-\braket{\bs a_j \bs a_\ell}$} are (linear) functionals of the two fields and  depend in an involved way on the Bogoliubov angle  of the transformation diagonalizing the Hamiltonian---see Appendix. 

In our specific protocol, significant simplifications occur close to the right edge of the lightcone. As detailed in Ref.~\cite{Maric2024long}, the relevant limit can be captured by a rescaled variable $z$ characterising the infinite time limit along the curve $z=z_{x,t}$, where $x=v_{\mathrm{max}}t+|v''(\bar p)|^{1/3} \Z_{x,t} t^{1/3}$,  $v(p)=\frac{d\varepsilon(p)}{dp}$ is the velocity of the excitation with momentum $p$, and $\bar p$ is the momentum associated with the fastest excitation ($v_{\mathrm{max}}=\max_p v(p)=2h$, $v''(\bar p)=-2h$). Analogously, momentum is described by a rescaled variable $q$ characterizing the infinite time limit along the curve $q=q_{p,t}$ with $p=\bar p+t^{-1/3} |v''(\bar p)|^{-1/3} q_{p,t} $. Note that the transformation $(x,p)\rightarrow(z,q)$ is canonical. Along the curve $(\Z_{x,t},q_{p,t})=(\Z,q)$ we find {a quickly oscillating} $\Psi_{x,t}(p)$ approaching $0$ 
{as $t^{-\frac{2}{3}}$} and the root density becoming proportional to the universal Fermi distribution that Refs~\cite{Bettelheim2011Universal,Dean2018} identified in a degenerate Fermi gas
\begin{equation}
\rho_{x,t}(p)=\rho_{\textsc{l}}(\bar p)\HS(2\Z+q^2)+O(t^{-1/3})\; .
\end{equation}
Here $\rho_\textsc{l}(p)$ is the root density 
{of} 
the left part of the state at the initial time and $\HS$ is a primitive of the Airy function, $
\HS(x)=\pi[\mathrm{Ai}(x)\mathrm{Gi}'(x)-\mathrm{Gi}(x)\mathrm{Ai}'(x)]
$, with $\mathrm{Gi}(x)$ one of the Scorer functions {(see e.g. Ref.~\cite{Antosiewicz1972})}. 
In the specific case of a thermal left reservoir,
$
    \rho_L(p)=\frac{1}{2\pi}\frac{1}{1+e^{\beta\varepsilon(p)}}
$.
Remarkably, the information about the left reservoir is encoded in a single parameter, $\rho_L(\bar p)$, confirming the irrelevance of most of the details of the reservoir. 

\paragraph{Order-parameter correlations.}

\begin{figure}[t]
    \includegraphics[width=0.95\linewidth]{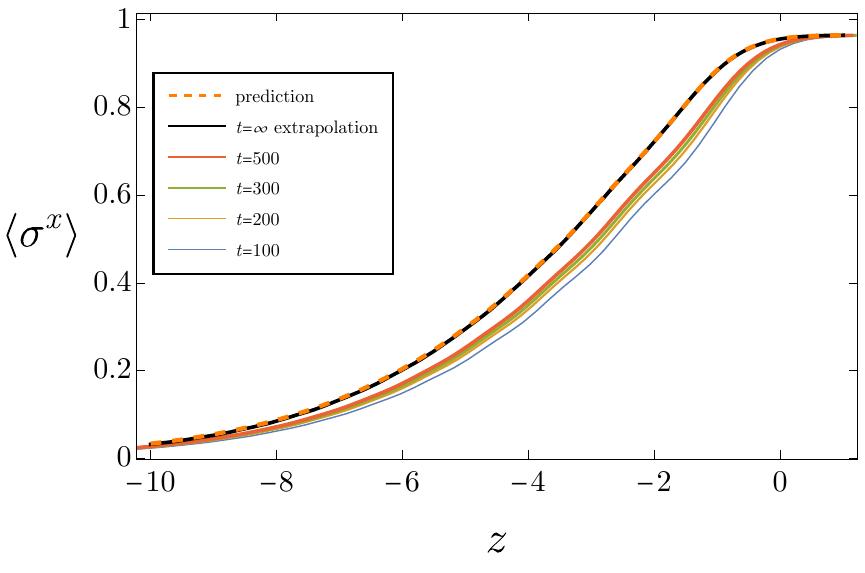}
    \caption{Magnetization $\braket{\sigma^x_j}$ near the right edge of the lightcone, as a function of $z=\frac{j-v_{\rm max}t}{|t v''(\bar p)|^{1/3}}$,  at different times $t$ for $h=0.5$ and $\beta=1$. The extrapolation {is obtained assuming corrections
    $a t^{-1/3}+bt^{-2/3}$ and fitting the coefficients $a,b$.}
    }
    \label{fig:mx}
\end{figure}

In the scaling limit we consider, the expectation value of an even---commuting with $\bs P$---local observable
approaches its ground state value; an odd---anticommuting with $\bs P$---local observable, $\bs O$, on the other hand, exhibits a nontrivial scaling limit:
\begin{equation}
\braket{\bs O_\ell(t)}= \braket{\bs O}_{\textsc{gs}} \mathcal M_\eta(z_{\ell,t})+O(t^{-1/3})
\end{equation}
where $\eta=-\log(1-4\pi\rho_{\textsc l}(\bar p))$, and, for a thermal reservoir,
$\eta=-\log [\tanh(\beta\varepsilon(\bar p)/2)]$, i.e., $\eta=-\log [\tanh(\beta\sqrt{1-h^2})]$.
Remarkably, the scaling function $\mathcal M_\eta$ does not depend on the specific odd observable and can be expressed as a Fredholm determinant {$\mathcal M_\eta(\Z)=\mathcal M_\eta(\Z,\infty)$, with}
\begin{equation}
{
\mathcal M_\eta(\Z_1,\Z_2)=\det\left|\mathrm I_{(\Z_1,\Z_2)}-(1-e^{-\eta}) \hat n_{(\Z_1,\Z_2)}\right|\; ,}
\end{equation}
where ${\rm I}_A(\Z_1,
\Z_2)=\delta(\Z_1-\Z_2)$ and $\hat n_{A}(\Z_1,\Z_2)$ is the Airy kernel
\begin{equation}
\hat n_{A}(\Z_1,\Z_2)=\\ \int_{-\infty}^\infty \frac{d q}{2\pi} \   \HS\left(\Z_1+\Z_2+q^2\right)e^{iq(\Z_1-\Z_2)}
\end{equation}
in the domain $z_1,z_2\in A$. We mention that in the infinite temperature limit ($\eta\rightarrow\infty$) $\mathcal M_\eta(z)$ approaches the (cumulative) GUE Tracy-Widom distribution $F_2(2^{\frac{1}{3}}z)$.  
More generally, we  refer the reader to Ref.~\cite{Tracy1993Level} for an expression of $\mathcal M_\eta$ in terms of the solution to a Painlev\'e II equation. 
Figure~\ref{fig:mx} reports a comparison between numerical data and prediction at different times for the local order parameter $\bs\sigma^x$.
The agreement is perfect. Note that $\mathcal M_\eta(\Z)$ approaches $0$ for large $-\Z$, confirming that the width of the interfacial region scales as $t^{1/3}$.
A similar result applies to the two-point functions of odd local observables $\bs O,\bs O'$
\begin{equation}
\braket{\bs O_\ell(t)\bs O'_n(t)}=\braket{\bs O_\ell\bs O'_n}_{\textsc{gs}}\mathcal M_\eta(\Z_{\ell,t},\Z_{n,t})+O(t^{-\frac{1}{3}})\, .
\end{equation}
As shown in~\cite{Claeys2018The}, just as $\mathcal M_\eta(\Z)$,  $\mathcal{M}_\eta(\Z_1,\Z_2)$ can also be expressed in terms of the solution to a system of ordinary differential equations.
Such asymptotic expressions are not compatible with clustering of correlations of odd operators in the interfacial region. In particular, the variance in a subsystem $A_t=(a^{-}_t,a^{+}_t)$ in the edge comoving frame, with $a^{\pm}_t= v_{\rm{max}}t+|v''(\bar p)|^{1/3}t^{1/3}\bar a^{\pm}$, scales as the square of the subsystem's length: denoting by $\braket{\bs O_{\ell}(t), \bs O_n(t)}_c$ the connected correlation $\braket{\bs O_{\ell}(t) \bs O_n(t)}-\braket{\bs O_{\ell}(t)}\braket{ \bs O_n(t)}$, the infinite time limit of the variance per unit length squared reads
\begin{equation}
\lim_{t\rightarrow\infty}\sum_{\ell,n\in A_t}\frac{\braket{\bs O_{\ell}(t), \bs O_n(t)}_c}{|A_t|^2\braket{\bs O}_{\textsc{gs}}^2}=\iint\limits_{\bar A}\frac{d^2 y}{|\bar A|^2} \mathcal M^c_\eta(y_1,y_2)
\end{equation}
with $\mathcal M^c_\eta(y_1,y_2) =\mathcal M_\eta(y_1,y_2)-\mathcal M_\eta(y_1)\mathcal M_\eta(y_2)$. 
This is strictly larger than zero for any finite $\bar A=(\bar a^-,\bar a^+)$.
The reader can find the details of the calculation and additional considerations in Ref.~\cite{Maric2024long}.

The scaling functions are universal also from another point of view: they are stable under localized perturbations at the initial time. On the other hand, {the scaling limit in which the perturbations occur at earlier but comparable times} are relevant, indeed Ref.~\cite{Maric2024long} shows \begin{multline}\braket{\bs O_\ell(t\cos\varphi),\bs O_n(t\sin\varphi)}_c\sim \\
\braket{\bs O}_{\textsc{gs}}^2\mathcal M^c_{\eta}(z_{\ell,t\cos\varphi},z_{n,t\sin\varphi},\varphi)
\end{multline}
for a scaling function {$\mathcal M_\eta^c(z_1,z_2,\varphi)$} that is strictly nonzero for $\frac{\pi}{4}\leq \varphi<\frac{\pi}{2}$.

We emphasize that these results are generalised straightforwardly to any noninteracting spin$-\frac{1}{2}$ chain described by a local one-site shift invariant Hamiltonian with a symmetry breaking ground state.

\paragraph{Skew information.}

\begin{figure}[t]
    \vspace{0.1cm}
    \includegraphics[width=0.95\linewidth]{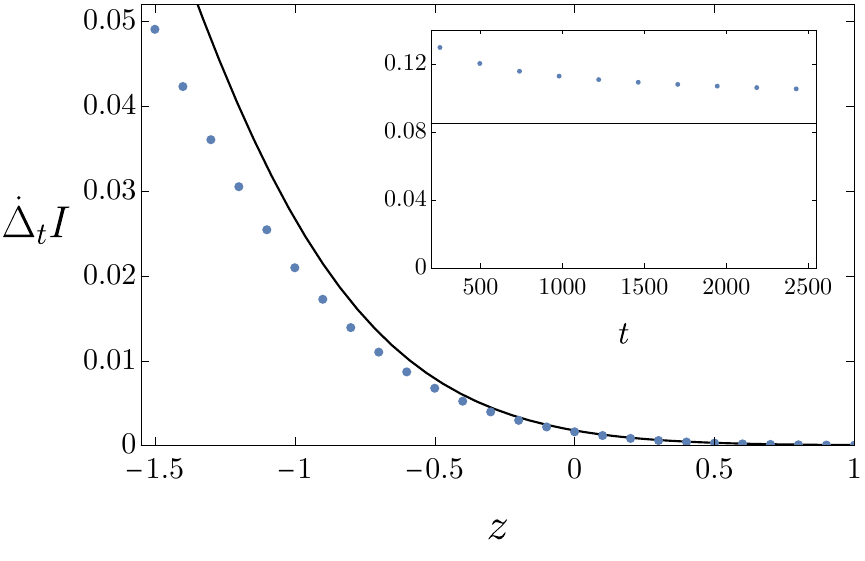}
    \caption{Increment of skew information $\dot \Delta_t I\equiv 4(I_{\rho_{A}}(\bs S^x)-I_{\tr_{\overline{A}}[\ket{\textsc{gs}}\bra{\textsc{gs}}]}(\bs S^x))/(t^{2/3}(m^x_{\textsc{gs}})^2)$ in half-infinite subsystems
    for $h=0.5$ and $\beta=1$. Main plot: $\dot \Delta_t I$ as a function of the position $z$ of the left boundary of $A=(z,\infty)$ for various times, compared with a semiclassical approximation. Inset: $\dot \Delta_t I$ as a function of time for $z=-2$; the horizontal line is the {extrapolated infinite-time limit, obtained as in Fig.~\ref{fig:mx}.}
    }
    \label{fig:skew}
\end{figure}

Recently, other nonequilibrium settings have been pointed out in which subsystems of quantum spin chains prepared in states with finite correlation lengths and time evolving under local Hamiltonians end up in states with low bipartite entanglement and no clustering properties~\cite{Bocini2023growing,Fagotti2024Quantum, Ferro2025Kicking}. Such phenomenology was accompanied by macroscopic quantumness. 
Roughly speaking, a macroscopic quantum state is a state in which there is some quantum behaviour that cannot be explained as an accumulation of microscopic quantum effects~\cite{Leggett1980Macroscopic}. A condition that is deemed sufficient for it concerns the behaviour of the quantum Fisher information $F_{\rho_A}(\bs O)$ of an extensive observable $\bs O=\sum_{\ell\in A}\bs O_\ell$ whose density $\bs O_\ell$ has support on a single site, such as $\bs S^x=\frac{1}{2}\sum_{\ell\in A}\bs\sigma_\ell^x$, where $\rho_A$ is the density matrix of some subsystem $A$. Specifically,
$F_{\rho_A}(\bs O)/(4|A|)$ is a lower bound to the size of the effective quantum space of $A$~\cite{Frowis2018Macroscopic} and it was also shown to be the convex roof of the variance of $\bs O$~\cite{Toth2013Extremal,Yu2013Quantum}, i.e., the minimal averaged variance over all possible decompositions of the density matrix in pure states. Since the Wigner-Yanase skew information $I_{\rho_A}(\bs O)=\tr[\rho_A\bs O^2]-\tr[\rho_A^{1/2}\bs O\rho_A^{1/2}\bs O]$ provides  a lower bound  to $F_{\rho_A}(\bs O)/4$, if $I_{\rho_A}(\bs O)$ scales as $|A|^2$ then $\rho_A$ is the density matrix of a macroscopic quantum state. Incidentally, $\frac{1}{2}I_{\rho_A}(\bs O)$ is also a lower bound to the quantum variance of $\bs O$, introduced in Ref.~\cite{Frerot2016Quantum}. 

We compute $I_{\rho_{A_t}}(\bs S^x)$ numerically, for $A_t$ in the edge comoving frame {with free fermion techniques~\cite{Maric2024long}}. We find 
$
\lim_{t\rightarrow\infty}t^{-2/3}I_{\rho_{A_t}}(\bs S^x)\neq 0
$.
Thus, subsystems  of size comparable with the interfacial region are in macroscopic quantum states. Figure~\ref{fig:skew} {compares}
numerical data
with a semiclassical approximation, detailed in Ref.~\cite{Maric2024long}. Although the quantum contribution to the variance is small in comparison to the classical part, $I_{\rho_{A_t}}(\bs S^x)$
witnesses quantum correlations over a distance of order 
$t^{1/3}$. 
\paragraph{Conclusion.}
We have investigated integrable quantum many-body systems in 1D, providing evidence---and rigorously proving in the transverse-field Ising chain---that the interfacial region formed by joining a symmetry-breaking ground state with a reservoir in a disordered phase exhibits full-range correlations, including a quantum contribution. 
However, we did not identify distinctive effects from interactions that preserve integrability, a point that warrants further investigation.
A significant open question remains whether this behavior could persist in the presence of integrability-breaking interactions.

\begin{acknowledgments}
This work was supported by the European Research Council under the Starting Grant No. 805252 LoCoMacro.
\end{acknowledgments}

\bibliography{references}

\appendix
\section{End Matter}
\paragraph{Appendix.}
We briefly review the connection between the fermionic correlation matrix and the fields $\rho$ and $\Psi$ characterising Gaussian states.
The Hamiltonian of the transverse-field Ising chain can be written as $\bs H=\sum_{j,\ell} \bs a_{j} \mathcal{H}_{j,\ell}\bs a_\ell/4$ for some Hermitian antisymmetric matrix $\mathcal{H}$. In the thermodynamic limit $\mathcal{H}$ is a block-Laurent operator generated by a $2$-by-$2$ symbol $h(p)$:
\begin{equation}
    \begin{pmatrix}
    \mathcal{H}_{2j-1,2\ell-1} & \mathcal{H}_{2j-1,2\ell} \\
      \mathcal{H}_{2j,2\ell-1} & \mathcal{H}_{2j,2\ell} 
    \end{pmatrix}=\int_{-\pi}^\pi \frac{dp}{2\pi} \ h(e^{ip})e^{ip(j-\ell)}
\end{equation}
with $h(e^{ip})=-2\sin (p) \sigma^x-2(h-\cos (p))\sigma^y$. The connection with the standard diagonalisation procedure involving a Bogoliubov transformation in the Fourier space is established by 
the following representation
\begin{equation}
    h(e^{ip})=\varepsilon(p) e^{-i\frac{\theta(p)}{2}\sigma^z}\sigma^y e^{i\frac{\theta(p)}{2}\sigma^z} \;, 
\end{equation}
where $\theta(p)$ is the Bogoliubov angle, given by $e^{i\theta(p)}=-(h-e^{ip})/|h-e^{ip}|$, and $\varepsilon(p)=2\sqrt{1+h^2-2h\cos p}$ is the energy of the quasiparticle excitation with momentum $p$.

It is customary to call a state Gaussian if the expectation value of every  local operator can be expressed in terms of solely  the correlation matrix $\Gamma_{j,\ell}=\delta_{j,\ell}-\braket{\bs a_j \bs a_\ell}$ through the Wick's theorem. 

For translationally invariant states, the correlation matrix can be represented as a block-Laurent operator with a $2$-by-$2$ symbol $\Gamma(e^{ip})$
\begin{equation}
    \begin{pmatrix}
    \Gamma_{2j-1,2\ell-1} & \Gamma_{2j-1,2\ell} \\
      \Gamma_{2j,2\ell-1} & \Gamma_{2j,2\ell} 
    \end{pmatrix}=\int_{-\pi}^\pi \frac{dp}{2\pi} \ \Gamma(e^{ip})e^{ip(j-\ell)}\, .
\end{equation}
This formalism can also be extended to inhomogeneous systems, where the correlation matrix is expressed in the following form
\begin{equation}\label{correlation matrix elements inhomogenous}
 \Gamma_{2j-2+i,2\ell-2+i'} = \int_{-\pi}^\pi \frac{dp}{2\pi} \ [\Gamma_{\frac{j+\ell}{2}}]_{i, i'} (e^{ip}) e^{ip(j-\ell)} \; ,
\end{equation}
with $i,i'\in\{1,2\}$. The symbol of the correlation matrix time evolves according to a Moyal dynamical equation, which is decoupled in the following representation
\begin{multline*}\label{correlation matrix symbol lqss}
  \Gamma_{x,t}(e^{ip})=e^{-i\frac{\theta(p)}{2}\sigma^z}\star\big[4\pi\rho_{x,t;o}(p) +(4\pi\rho_{x,t;e}(p)-1)\sigma^y  \\ +4\pi\Psi_{x,t;R}(p)\sigma^z-4\pi\Psi_{x,t;I}(p)\sigma^x \big]\star e^{i\frac{\theta(p)}{2}\sigma^z}  \; ,
\end{multline*}
where $\rho_{x,t}(p)$ is the root density, $\rho_{e,o}=(\rho(p)\pm \rho(-p))/2$ are its even and odd part, respectively, and $\Psi_{x,t}(p)=\Psi_R(p)+i\Psi_I(p)$ is an auxiliary field, which is odd under $p\rightarrow -p$. The operation $\star$ is the Moyal product, which is formally defined as follows
\begin{multline*}
    (f\star g)(x,p)
    =\sum_{m,n\in\mathbb{Z}} e^{i(m+n)p} \iint_{-\pi}^{\pi} \frac{d^2q}{(2\pi)^2} \\
    e^{-i(nq_1+mq_2)}f(x-\tfrac{m}{2},q_1)g(x+\tfrac{n}{2},q_2) \ .   
\end{multline*}
Finally, the root density and the auxiliary field satisfy the dynamical equations reported in \eqref{eq:dyneq}.

In the limit of low inhomogeneity the Moyal equation can be expanded in the order of space derivatives. Keeping the first two non-zero orders gives the third order generalized hydrodynamic equation
\begin{equation}\label{third order generalized hydrodynamics}
\partial_t\rho_{x,t}^{(3)}(p)+v(p)\partial_x\rho^{(3)}_{x,t}(p)=\tfrac{v''(p)}{24}\partial_x^3\rho^{(3)}_{x,t}(p)\ ,
\end{equation}
where $v(p)=\frac{d\varepsilon(p)}{dp}$ is the velocity of the quasiparticle excitation with momentum $p${, and the superscript $(3)$ is a reminder that $\rho$ is equal to $\rho^{(3)}$ only up to subleading contributions---in the limit of infinite time---coming from spatial derivatives higher than the third}.
The third-order correction  is particularly important close to the  edge of the lightcone, where it gives a leading contribution.
In a bipartitioning protocol where one side of the system, e.g., the right-hand side, is initially prepared in a stationary state, the auxiliary field {is instead characterized by rapid oscillations and an amplitude that approaches zero as $t^{-\frac{1}{2}}$
along any ray with a finite slope strictly within the lightcone, and as  $t^{-2/3}$ at the edge of the lightcone~\cite{Maric2024long}}.
\end{document}